\documentclass[reprint,
superscriptaddress,
 amsmath,amssymb,
 aps,
]{revtex4-2}

\usepackage{graphicx}
\usepackage{dcolumn}
\usepackage{bm}
\usepackage{gensymb}
\usepackage{textgreek}
\usepackage{listings}
\usepackage{pythonhighlight}
\usepackage{mathtools}

\begin{document}


\title{A Machine Learning Model for Predicting Progressive Crack Extension based on DCPD Fatigue Data}

\author{Jacob Keesler-Evans}
\affiliation{Department of Mechanical \& Aerospace Engineering, West Virginia University}
\author{Ansan Pokharel}
\affiliation{Department of Mechanical \& Aerospace Engineering, West Virginia University}
\author{Robert Tempke}
\affiliation{Department of Mechanical \& Aerospace Engineering, West Virginia University}
\author{Terence Musho}
\email[corresponding author:]{tdmusho@mail.wvu.edu}
\affiliation{Department of Mechanical \& Aerospace Engineering, West Virginia University}
\begin{abstract}
Time history data collected from a Direct Current Potential Drop (DCPD) fatigue experiment at a range of temperatures was used to train a Bidirectional Long-Short Term Memory Neural Network (BiLSTM) model. The model was trained on high sampling rate experimental data from crack initiation up through the Paris regime. The BiLSTM model was able to predict the progressive crack extension at intermediate temperatures and stress intensities. The model was able to reproduce crack jumps and overall crack progression. The BiLSTM model demonstrated the potential to be used as a tool for future investigation into fundamental mechanisms such as high-temperature oxidation and new damage models.

\end{abstract}

\maketitle


\section{Introduction:}
High-temperature fatigue is an extremely complicated process~\cite{totemeier2007creep,kawagoishi2000fatigue} that requires a multi-physics perspective~\cite{gustafsson2013modelling} for new advancements of new and existing alloys can be realized. One of the difficulties in understanding fatigue is understanding the relationship between material properties and material performance. Often the root of these difficulties is their interdependence and extremely non-linearity~\cite{stephens2000metal}. Temperature, loading conditions, and environment on top of microstructure influence the material performance~\cite{gustafsson2011influence}. While there has been steady progress over the last century on characterizing low-cycle fatigue properties, the ability to fully predict high-temperature fatigue is still in its infancy~\cite{pook2007metal,ding2007brief,chan2010roles,biallas2007situ}. One tool that is gaining traction is the application of machine learning (ML) techniques~\cite{kamble2021machine,zhan2021novel,luo2021pore,sanchez2021machine,liu2020predicting}. Deep learning and other such ML tools have begun to be utilized for improved life predictions, which have shown promise~\cite{ZHANG2021106236, mlfatigue}. While ML cannot currently solve the whole problem it provides a method of pattern recognition and correlation by utilizing massive amounts of data. This allows information present in individual experimental trials to be investigated, instead of relying on statistical trends taken over multiple trials. The following research applies the concept of ML modeling by focusing on data collected from the direct current potential drop (DCPD) experimental method to train an ML model. This DCPD method is a common experimental technique defined by an ASTM standard~\cite{astm2018dcpd} for fatigue that can, as done in this study, provide high-resolution temporal data at the crack tip. To limit the scope of this research, the aim is to demonstrate that a machine learning modeling can reproduce features from the experiment and provide equivalent interpolate and extrapolate experimental data, thus providing a computational tool for future fundamental studies and new damage model development.

\begin{figure}[!ht]
    \centering
    \includegraphics[width=0.9\textwidth]{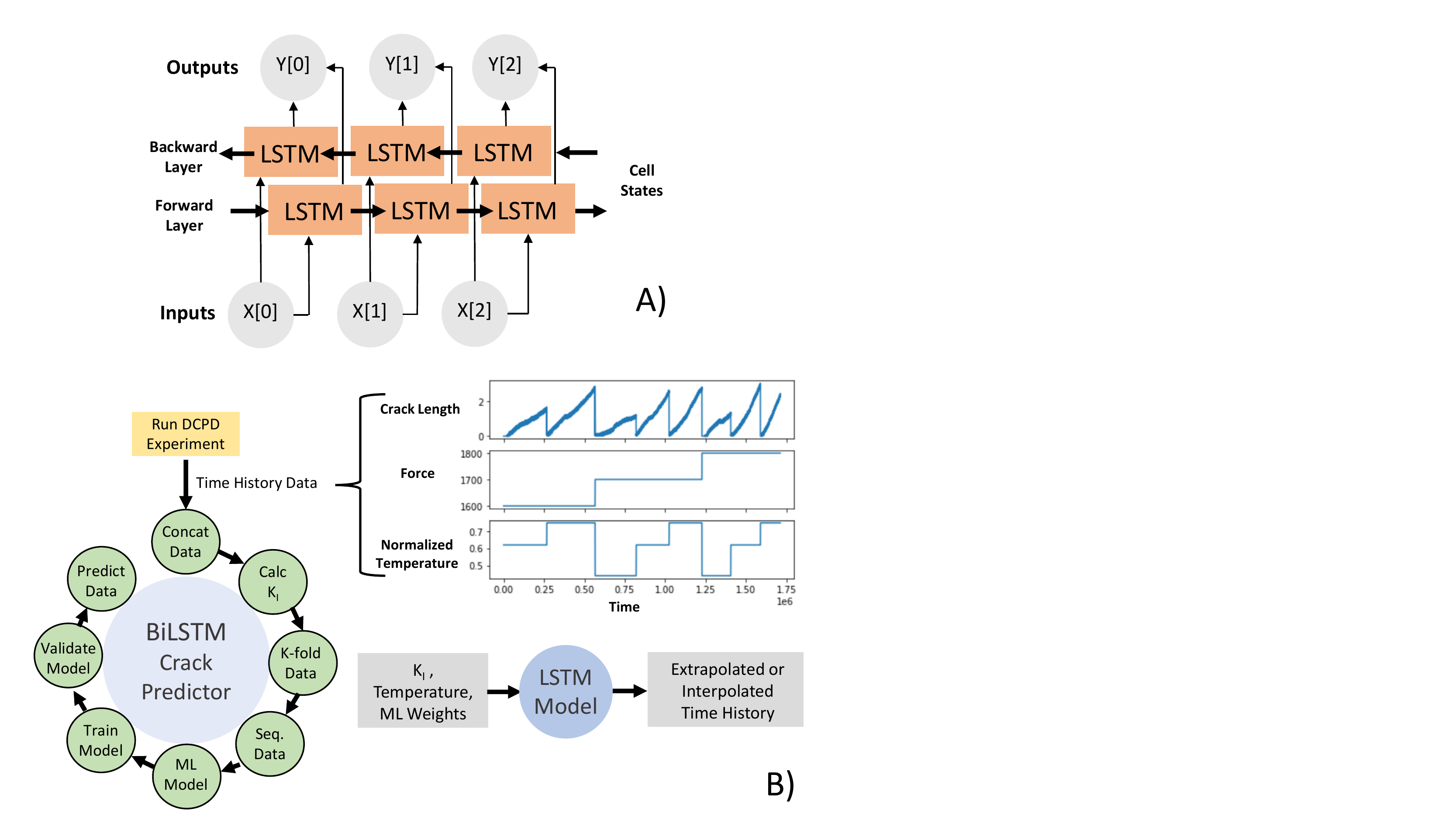}
    \caption{\label{fig:cycle}Illustration of the machine learning process applied to predict crack propagation time histories at both temperature and pressure. Input to the model is stress intensity and temperature. Sub-figure A is a high-level illustration of the BiLSTM utilized. Sub-figure B is an illustration of the training process and prediction process.}
\end{figure}

There have been several reviews of the current state of fatigue modeling using machine learning modeling techniques within the linear Paris regime~\cite{kamble2021machine,wang2017comparison,hu2020bayesian}. Wang et al. discussed the nonlinear relationship that has been observed experimentally between fatigue crack growth rate and stress intensity factor. To get a good agreement, however, they had to utilize a multistage machine learning technique to predict crack growth in superalloys within the Paris region~\cite{paris1999service}. This same problem and solution of crack propagation and machine learning solutions have been applied to other polycrystalline alloys~\cite{rovinelli2018using}. Rovinelli et al. used a machine learning technique to predict the direction and speed of small crack advancement in polycrystalline materials. Their study focuses on utilizing machine learning to understand the microstructural variables that influence crack propagation. In a similar study of hyper-elastic materials, Joeun Choi et al. used deep learning to get an average error of 14.3\% for crack growth predictions in polychloroprene rubber reinforced with tungsten nano-particles~\cite{choi2021multiaxial}. An approach by Hu et. al. used a similar machine learning approach to optimize the parameter of a Bayesian model~\cite{hu2020bayesian}. It is found that many of these studies rely on semi-empirical models that are parameterized using a machine learning approach. While this approach is valid, it integrates a level of supervision into the ML model. This approach becomes more curve fitting or regression than feature recognition. The latter, also the interest for this study, is more akin to image processing that leads a path towards unsupervised or artificial intelligence (AI) modeling in the absence of prior knowledge. While the unsupervised approach is often desired it is found to depend on the desired modeling outcome and the state-of-the-art of machine learning technology, which governs the most appropriate approach.

The following study utilizes a Bidirectional Long-Short Term Memory Neural Networks (BiLSTM) in the capturing of general fatigue crack length prediction and the prediction of individual events that occur in crack length data. Utilizing high-resolution temporal and spatial data that properly captures crack growth characteristics, the BiLSTM can be used to generate equivalent experimental fatigue crack growth data in place of running many different, potentially costly, laboratory experiments. The BiLSTM applied is an unsupervised learning approach, where the ML is permitted to recognize any feature within the time series.

The Long-Short Term Memory Neural Networks (LSTM), specifically the bidirectional variation shown in Figure~\ref{fig:cycle}A, was chosen for its long term data persistence, its ability to reach better performance, and its ability to extract additional features that other neural networks, including the standard LSTM, which was found to have difficulty capturing time history features.\cite{LSTMvsBLSTM, StockBLSTM}. Figure~\ref{fig:cycle}B is an illustration of the overall process developed and studied for this research. The machine learning model is trained on high sampling rate data from an experimental direct current potential drop measurement of in-situ monitoring of a crack extension. The goal of using this approach is the ability to reproduce experimentally equivalent data at intermediate test conditions, be it, stress intensity or temperature. One of the important pieces of information contained within the DCPD data is the rate of oxidation and how it influences the crack jump statistics. It has been demonstrated that the time history data from the DCPD experiment for an individual sample is a non-random process~\cite{ansan}. More specifically, it can be demonstrated that crack events early in the time history influence what happens later in the experiment. Currently, this information is often lost because statistical averages are computed over sample sets, and the sampling rate when calculating attributes, such as the change in the crack, is often taken over large time steps (minutes). While machine learning models do not provide a fundamental understanding of the physics, it does provide a tool for analyzing large amounts of data. Additionally, the machine learning approach provides a method of pattern recognition and correlating inputs to outputs that are beyond the capabilities of an individual.

One of the main applications of the developed model is the ability to train the machine learning model on a select number of experimental trials and then use the machine learning model to not only interpolate the data but also extrapolate the data. This provides a more accurate way of correlating inputs to output in the absence of conducting more experiments, which are time-consuming (12-16hr per sample). Contain within the model are features that rely on the underlying physics, which provides a prediction of new physics such as oxidation at high temperatures. One specific feature of interest is the notion of crack jumps, which is hypothesized to depend on prior events in time history.

The developed model in this study relies on time history data from fatigue data that was collected in a previous study~\cite{ansan} for the study of a heat-treated Inconel 718 (IN718) material. IN718 is a nickel-based superalloy that is specifically designed to withstand operating temperature upward of 700\degree C. This material is precipitated hardened, meaning the microstructural features are critical to the material performance at elevated temperatures. This material has excellent fatigue properties, along with oxidation resistance at high temperatures. A common application of this material is within gas turbine engines for the component that secures the turbine blades to the rotating shaft, known as a turbine disk. The DCPD experiment was run at multiple loading conditions (initial stress intensities) and temperatures. This data was used to train the ML model. An illustration of the overall process is shown in Figure~\ref{fig:cycle}. The ML model is trained based on temperature and stress intensity and predicts the crack extension. After training the ML model, the model can predict at intermediate temperature and stress intensities and can extrapolate to high-temperature values.

\subsection{LSTM Theory:}
The research conducted in this study relies on the LSTM framework implemented in TensorFlow, and as such, the general theory behind the LSTM is well documented. While it is beyond the scope of this article to detail all the mathematical details, it is necessary to provide a high-level explanation so model adjustments can be discussed. The Long Short-Term Memory Neural Network is a type of  Recurrent Neural Network (RNN) architecture that retains beneficial properties while improving general network functionality. The most beneficial property of the RNN, and by extension the LSTM, is the persistence of data through the network. This persistence is achieved through the use of a cell state, which flows through the cells in the network. This means that each cell can access and edit the cell state, thus adding additional information or removing irrelevant information.

The difference between a traditional RNN and the LSTM is the time dependency limitations, with LSTM's able to bridge amounts of time that exceed that of the RNN~\cite{hochreiter1997long}. In the case of the RNN, long-term dependencies could only be handled with the meticulous selection of hyper-parameters~\cite{DBLP:journals/corr/abs-1803-00144}. The issue then becomes with the amount of time and computational resources that would be required to select the required hyper-parameters for the RNN. This is the opposite case of the LSTM which can handle large time gaps and will perform well over a range of hyper-parameters~\cite{hochreiter1997long}.

The ability of the LSTM to handle long-term dependencies and large time gaps is inherent to the system architecture. Whereas initial forms of the LSTM could handle long-term dependencies, it still had the issue of gradient either vanishing or diverging. In these cases, the gradient is driven so low incoming errors are blocked, or so high that it will cease functioning as a memory cell~\cite{vangrad}. This means that with that when the gradient of the loss function gets so low that training becomes increasingly difficult, and with exploding the build-up of error will lead to an unstable network. Put even simpler, either error gets extremely large and leads to total degradation of the network or shrinks, which causes a decrease in learning ability. 

In order to deal with the gradient issues, forget gates were implemented. These forget gates serve as either an immediate and total, or slow and gradual reset to memory blocks. The forget gate can assist in gradient issues and memory issues by freeing up additional memory. However, memory isn't removed at random, only out of date and useless information is removed from memory which in turn deals with the exploding and shrinking of error gradients\cite{vangrad}. 

With the relevant portions of a traditional LSTM discussed, it should be noted that various variations of the LSTM architecture exist. Each has its unique use, however most importantly for this study is the bidirectional LSTM. A bidirectional RNN is an RNN that processes the data both forwards and backward, with one cell state flowing forward which represents data flowing with positive time. The other state flows backward and is considered to have negative time, see Figure~\ref{fig:cycle}A. When both of these cell states pass through a network, this allows the network to use both the past data from the time positive cell state and future data from the time negative cell state~\cite{Bidirectional}. A bidirectional LSTM, henceforth referred to as BiLSTM, is just an LSTM as discussed prior, but with both the time negative and positive cell states. This allows for the LSTM to use long-term contextual data from both the past and future to drive down the error and achieve better results. In general, the training methods for the application of the LSTM are the same as with the regular LSTM, the only difference occurs when back-propagation is used. In this context, the bounds need special treatment since either forward or backward state inputs are unknown at the time\cite{Bidirectional}.

\section{Method:}
\subsection{Computational Method:}
The BiLSTM model was constructed using TensorFlow release 2.5. The model was executed on Google's Colab system. The hardware was selected to be GPU-based using TensorFlow-GPU libraries. The GPU hardware did vary on the Colab system but varied between Tesla P100, Tesla V100, and Tesla P4. The solution did not vary over different hardware, only execution time. The precision was the data was allowed to be mixed precision when running on GPUs.

The model was written in Python3 with a Keras high-level interface over the TensorFlow back-end. A series of normalization techniques were used to scale the data. As seen in Figure~\ref{fig:cycle}B the time history data was initially concatenated. Testing was done to ensure that the model accuracy was independent of concatenation order through the initial randomization of the order in which the files were read in. It was found that there was no change in model performance between the randomized order of samples, and the non-randomized order. This lack of change is due to time-series properties of the individual samples still retaining their proper order no mater the order of reading. Since the data was sampled at regular intervals it was no necessary to carry along the time. The data from this step forward was treated as steps or cycle numbers. The stress intensity was calculated based on Equation~\ref{eq:1}, which is based on a single-edge crack in tension. The stress intensity incorporates both the force and the crack length. The temperature and stress intensity are used to predict the crack length. The data was trained on 90\% of the data and the remaining 10\% was used for validation. A K-fold method was used to fold the validation data back onto the data. This was necessary because the experimental data were concatenated. The data was then sequenced using a method of inputting three steps and predicting on the fourth. This sequencing technique was used for all the steps. The BiLSTM model was determined to be fairly small with only 100 neurons, batch size 256, and dropout of 20\%. A snippet of the Keras model that was used is listed as the following.

\begin{python}
	opt = Adam(learning_rate = 0.0005)
	model = Sequential()
	model.add(Bidirectional(LSTM(100,
		activation='tanh'
		,recurrent_activation='sigmoid')))
	model.add(Dropout(0.2))
	model.add(Dense(1))
	model.compile(optimizer=opt, loss='mse')
\end{python}

Much more complex networks were tried during this study but it was found that they had a larger mean square error when compared to the above network. Deep networks with up to five layers with dropout between each layer were not as well-performing. It was noted in these more complex networks that the models either had a significant offset or had difficulty predicting late in the time series. With the simple network, the activation function and the recurrent activation functions were selected to use the tanh and sigmoid activation functions. As will be discussed in the results section, because the network is fairly simple it will require the input to contain high-frequency features to produce high-frequency features in the output.

\subsection{Experimental Method:}
The experiments were run in a material testing system MTS810 hydraulic load frame and a direct current potential drop method was employed in-situ to capture the time history data. The testing followed the ASTM compact specimen standard testing procedure\cite{astm2015standard}. A custom test rectangular specimens of dimension 76.2mm x 36.63mm x 0.46 mm with two holes for mounting were prepared. In order to capture the crack initiation event and later stages of crack growth, the samples were initiated with a 10mm initial crack created using an electro-discharge machining (EDM) technique.

A nickel-based superalloy (IN718) was chosen for this study based on previous experimental studies and availability of data~\cite{ansan}. IN718 has a high operating range, this range spans from -423\degree F up to 1300\degree F. The favorable performance of IN718 is due to its complex alloy of 15+ elements, along with the production precipitate hardening the alloy\cite{Inconel718}.

IN718 samples were prepared from two different heat treatment conditions. The first condition included as-received annealed samples while the second condition comprised of fully heat-treated samples, prepared following the ASTM standard\cite{astm2018standard}. The heat-treated sample had a Rockwell-C hardness of 41~\cite{ansan}. The annealed sample had a Rockwell-C hardness of 21.1. The data discussed within will only pertain to the heat-treated sample.

The crack length measurements were taken using the DCPD method. The DCPD technique uses the size of a growing crack in conjunction with Gauss's Law to measure the drop in potential between two probes placed on either side of the crack\cite{DCPDacc}. As the experiment is continuously ran, the crack length will increase and the potential field will change between the voltage probes. The crack length was correlated with the potential at the probes using a calibration function. The calibration function was derived using the geometric properties of the test specimen and the Johnson equations~\cite{johnson1965calibrating}. DCPD was chosen for its high precision measurements and ability to capture dynamics at the crack front. Moreover, the DCPD can be ran at a high sampling rate, for this study 512 samples/sec, to provide high spatial and temporal resolution of the crack evolution. During this experiment, we capture both crack initiation (region I or stage I) and crack propagation (region II or stage II). This second region is well characterized as the Paris regime~\cite{paris1999service}. The experiment is stopped just prior to region III, which leads up to a failure of the sample. 

In order to run the fatigue experiment, the test specimen was run through a loading cycle which consisted of two parts. The first part consisted of 10 fast triangular oscillations for 30 seconds that were run at a given minimum and maximum load (R=0.15). In the case of this experiment, the peak loads were chosen to be 1600N, 1700N, and 1800N. Following the oscillations, there was a 100-second hold period in which the test specimen was held at the peak load\cite{ansan}. The hold cycle is the part of the interest as this is when the crack will jump. The different peak loads define the initial stress intensity and subsequent stress intensity. The experiment was carried out at a series of temperatures by heating the samples and simultaneously loading the sample. It is important to note that the crack dynamics are strongly correlated to the temperature and the oxidation environment. All of the samples were run in an atmospheric environment.

As previously stated, the data collection occurred at a sampling rate of 512Hz. This was done in an attempt to capture all of the crack jumps since they occur at such small length scales and short time duration. Post-processing of data was done to extract the crack jumps only during the 100-second hold cycles.

\section{Discussion and Results:}
The IN718 results discussed within are for the ASTM specification heat-treated condition for IN718. The heat-treatment results in a precipitate hardened state with the $\gamma$" phase being the dominant strengthening phase in IN718. The ML model is able to predict accurately the annealed condition but this is a less relevant condition as the material performance is degraded in the annealed condition.

As seen in Figure~\ref{fig:cycle}, the data provided from the experiment is time history data where the inputs are the temperature, peak load, and crack length. The peak load is converted to a stress intensity because it is well understood that the crack propagation is correlated to the stress intensity. The stress intensity is enhanced by the crack tip, the length of the tip, and the geometry of the sample. The stress intensity equation for a singe-edge crack in tension takes the following form,

\begin{equation} \label{eq:1}
\begin{split}
    K_I=&\sigma\sqrt{\pi a}[1.122 - 0.231(\frac{a}{W}) + 10.55(\frac{a}{W})^2 \\
        &- 21.71(\frac{a}{W})^3 + 30.382(\frac{a}{W})^4].
\end{split}
\end{equation}
Where $\sigma$ represents the uniform stress, $a$ represents the crack length from the edge, and $W$ represents the width of the sample in use. The polynomial on the right side of the equation is a geometric enhancement factor~\cite{ansan}. The crack length is based on the initial crack length and the additional crack extension as the crack propagates. As is noted from Equation~\ref{eq:1} as the crack increases the stress intensity increases. Additionally, the peak load sets the initial crack intensity. 

The model is trained on the stress intensity, temperature and predicts the crack length. As noted in Equation~\ref{eq:1} the stress intensity combines the peak force and crack length. This is an important aspect of this modeling approach. If the crack intensity is known and the temperature the model will predict the change in crack length. The model is trained on the notion of the Paris regime where the change in crack length per cycle ($\delta a/\delta n$) can be related to the change in stress intensity ($\delta K$). However, this relation is typically linear in the Paris region, wherein in this case, the machine learning model can capture the non-linearity of region 1 up to region 2. This notion of the crack length is directly correlated to the input and output is the reason why the bidirectional LSTM model was selected for this study. Overall, this input structure provides a method of interpolating and extrapolating to other stress intensities, as will be discussed below. Moreover, this workflow provides a method of training damage models in the future.

It is important to point out that the model is being trained on experimental data in the absence of filtering the data. The machine learning model is be trained so that it learns both the noise features in the data and the overall trend of the crack dynamics from initiation up through propagation as a function of temperature. It should be noted that it will be necessary to supply the model with similar noise inputs to produce noise features. This is because the model input of stress intensity contains noise. The word noise in this context does pertain to random noise originating from the DCPC but it also contains information about the underlying physics at the end of the tip, that is not well defined. This may include oxidation of the crack tip, which is known to influence the crack dynamics. The model is capturing these features as they are reproducible in the output as will be seen below.

\begin{figure}[!ht]
    \centering
    \includegraphics[width=0.48\textwidth]{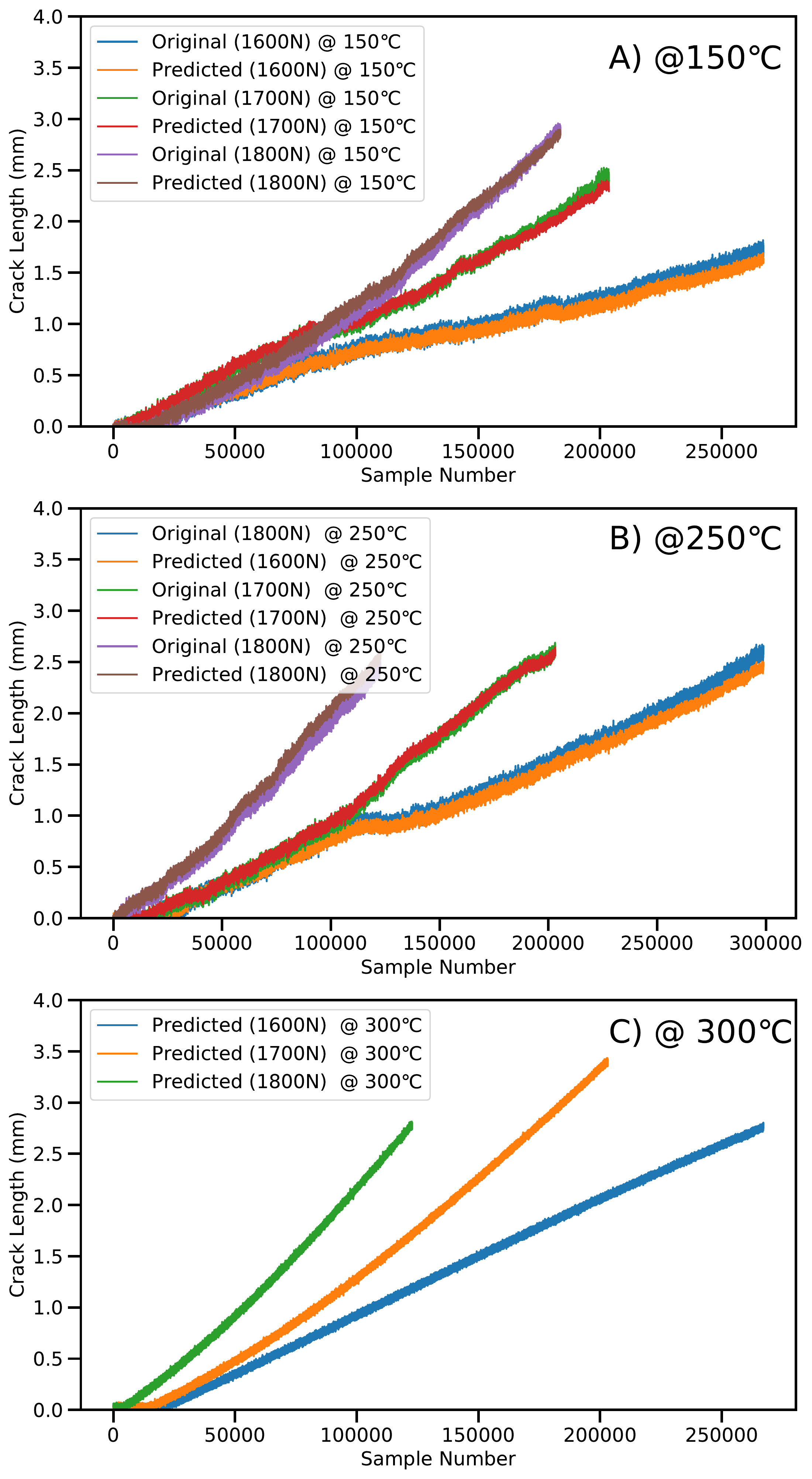}
    \caption{\label{fig:CL}A plot of the crack length versus cycle number. Sub-figure A is a plot of the experimental and model-predicted values at 150\degree C. Sub-figure B is experimental and predicted at 250\degree C. Sub-figure C is the extrapolated model value at 300\degree C.}
\end{figure}

To validate the model, the weights were saved from the training of the model, as illustrated in the left portion of Figure~\ref{fig:cycle}B. Model validation was performed primarily on two different types of data. The first type, was the same data it was trained on. Using this training data for validation allows for the determination of if the model adequately learned the underlying data. A poor fit to existing data would indicate issues with the model. Alternatively, interpolated data was also used for validation. Using known experimental data of a given force, the data of the different temperatures were fit to a two dimensional polynomial. This polynomial was used to generate interpolated and extrapolated data to validate the model on. Figure~\ref{fig:CL} is a plot of both the original data and the model predicted data. Figure~\ref{fig:CL}A is at 150\degree C for the three peak loading conditions. Figure~\ref{fig:CL}B is a similar figure for the three peak loads at 250\degree C. This data is interpolated using the aforementioned two dimensional polynomial. Both A and B are in good agreement with the original data that was used for training. At both temperatures, the model prediction is able to pick up the noise features and the overall macro-trends. These macro-trends are two-fold. The first being that the crack length increases with the sample number. It can be noted in these plots there is a change in the slope when the crack dynamics transition from the region I to region II. The second macro-trend is that with increase peak load there the slope or rate of crack propagation is different. 

Figure~\ref{fig:CL}C is an extrapolated temperature case where a extrapolated value of stress intensity and a elevated temperature of 300\degree C is fed into the model and the crack length is predicted. As mentioned previously the model is trained on the actual experimental data that contained noise. Again this noise is a result of both random noise from the DCPD experimental measurement and some underlying physics at the crack tip (oxidation, grain boundaries, etc). As seen in Figure~\ref{fig:CL}C there is some noise in the predicted data. This is a result of inputting noise with the input stress intensity that is proportional to the random noise of the DCPD measurement. The random noise of the DCPD measurement was quantified in a previous study~\cite{ansan} as standard deviation of 0.025. In the absence of this input noise in the stress intensity the model will return a smooth curve. This is a direct observation the machine learning model is prediction the output features based on the input features. This is a significant finding as this is exactly how the experiment would evolve in person when conducted. Any change in the stress intensity at the crack tip would result in a change in the crack length.

Another feature that is captured by the model crack initiation time or lift-off time. This can be seen in Figure~\ref{fig:CL}C. When the peak stress or initial stress intensity is lower the time it takes for the crack to lift off is longer. This is directly related to the geometry of the initial crack and is often not captured in DCPD type experiments because of a crack initiation procedure. However, in this experiment, we were interested in the initiation process and allowing the machine learning model to capture the feature of this initiation process. Again, this will depend on the input stress intensity but lift-off time and slope in this region will be determined by the machine learning model.

\begin{figure}[!ht]
    \centering
    \includegraphics[width=0.48\textwidth]{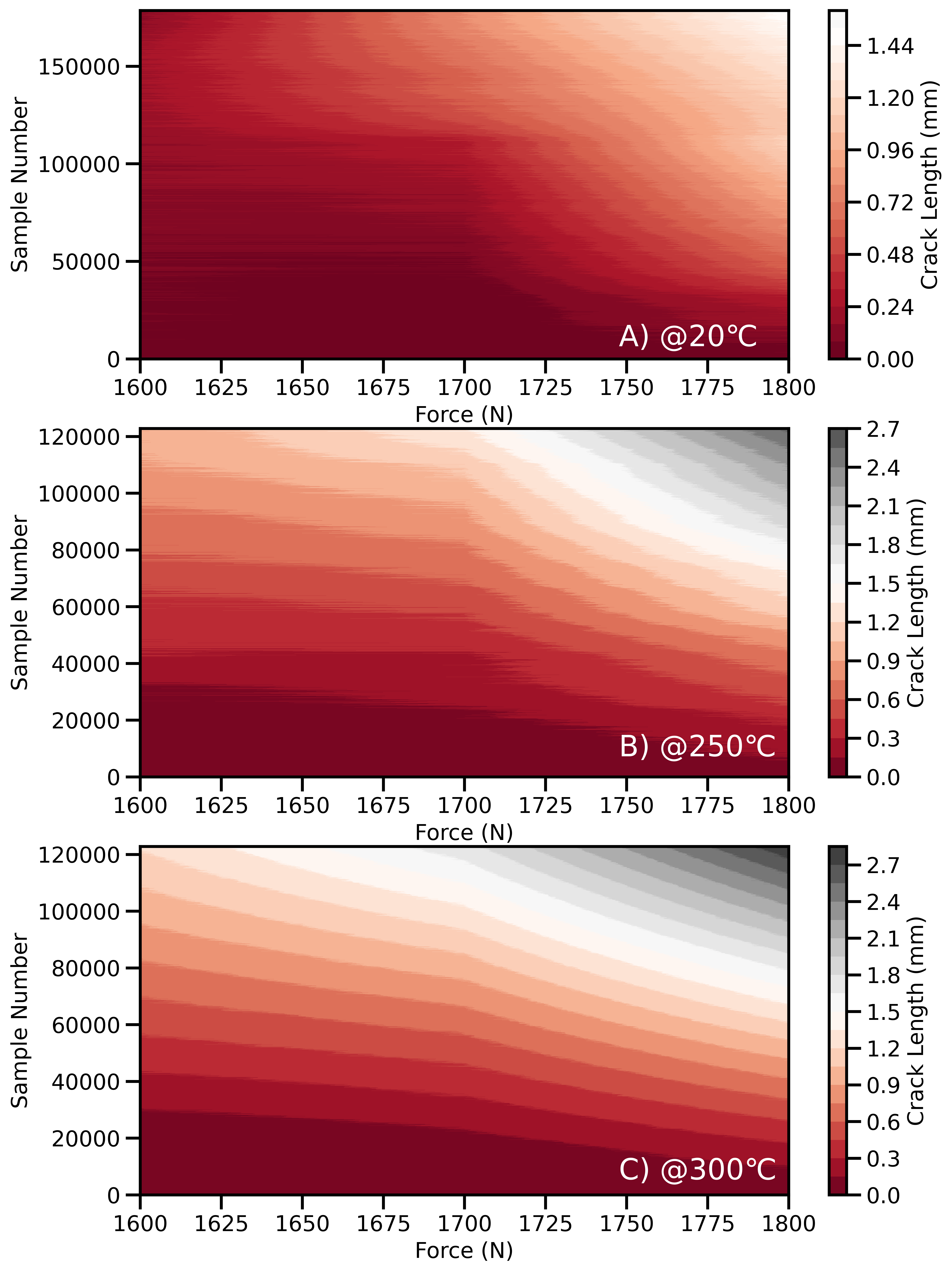}
    \caption{\label{fig:heatmap}Contour plot of the crack length versus sample number and load. The color contour corresponds to the crack length. Subfigure A is at room temperature, Subfigure B is at 250\degree C, and Subfigure C is at an extrapolated temperature of 300\degree C. Notice when temperature increases crack length increases }
\end{figure}

Additionally, the machine learning model was tested to see how well it can predict intermediate values of peak loads or intermediate initial stress intensities. Figure~\ref{fig:heatmap} is the contour plot of sample number versus peak force. The contour colors correspond to the crack length. Figure~\ref{fig:heatmap}A is room temperature, B is 250\degree C, and C is at an extrapolated temperature of 300\degree C. Similar trends to Figure~\ref{fig:CL} are seen in Figure~\ref{fig:heatmap}. It is expected that the force relationship with crack length will be continuous. Essentially every possible force value between 1600N and 1700N will have a solution. A similar argument can be made for the temperature however care must be taken extrapolating to a high temperature as the crack dynamics may transition. It is well known that fatigue transitions from transcrystalline to intercrystalline as the temperature increases~\cite{WANG19933}. However, it is hypothesized that this machine learning approach should be able to capture this transition if experimental data is available through this transition.

\begin{figure}[!ht]
    \centering
    \includegraphics[width=.48\textwidth]{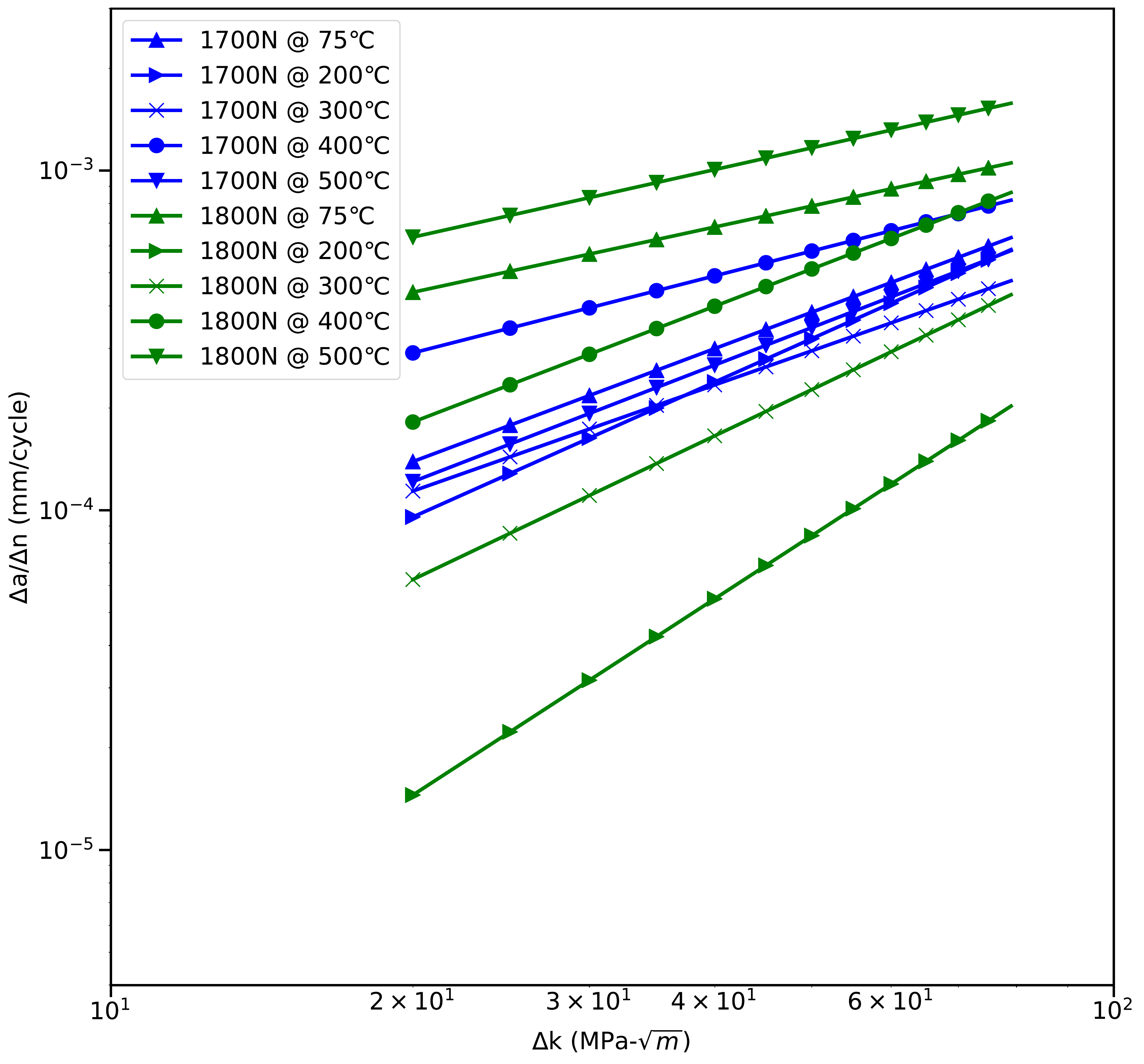}
    \caption{\label{fig:dadn} Plot of crack length change over number of cycles against the number of cycles on a logarithmic scale based on machine learning produced interpolated and extrapolated results. }
\end{figure}

To further prove the ability of the model to predict the crack dynamics at elevated temperatures a plot of the change in crack length per cycle versus change in stress intensity. The results are shown for two of the peak load condition in Figure~\ref{fig:dadn}. Several interpolated and extrapolated temperatures are displayed that range in temperature from 75\degree C to 500\degree C. The values are calculated based on the mean values for change in crack length (y-axis) and stress intensity (x-axis). The range of the data has been selected to display only the Paris or region II crack propagation. It is noted from Figure~\ref{fig:dadn} that two important concepts are reproduced by the machine learning model. First if that there is a shift in the positive y-direction for increasing temperature. Again, one should be careful with extrapolating across and transcrystalline to the intercrysalline transition. The slope of the curves. The sloped should be constant for a given peak load. This slope will depend if the data is sampled near the tails of the Paris regime, which is the case for the 1700N 75\degree C case. Again, the ability to interpolate and extrapolate values at intermediate stress intensities and temperature is critical to developing a damage model with increasing accuracy.

The potential overfitting of the model was taken into account during the training process and was accounted for though the use of 20\% dropout in the model. Due to the small size of the model and due to the fact that some of the features from the noise could hold important information regarding crack propagation, as such this data should not be lost. To assist in testing for overfitting, the interpolated data was used as new data, if there was a large disagreement with the two dimensional polynomial values and the predicted values, overfitting of the model was certain. However, this was not the case. As can be seen in Figure~\ref{fig:CL} B, this is not the case. Both sets of data are largely in agreement, leading to the belief that there was no overfitting of the model. 

In order to judge the accuracy of the model, the mean absolute error (MAE) was used. Since the MAE represents the absolute difference between the value of the predicted value and true value, the MAE gives a good idea of the average error. As such, this leads to a generalized idea as to the accuracy of the model. For this model, the MAE was found using both the training data and the k-folded validation data. Upon the calculation of the MAE of the training data, the value was found to be approximately 0.03899. This indicates a low deviation from the training data on average. Similarly, the MAE of the testing data was slightly higher, as expected, having a value of 0.0456. Again, with the training data, there is a low deviation of the predicted value from the truth. This can be seen in Figure~\ref{fig:CL}, where a large majority of the predicted cases intersect with the truth values, with minor deviation. 

\section{Conclusion:}
A machine learning model has been investigated that takes time history data collected from DCPD to predict crack length evolution based on stress intensity and temperature. A bidirectional LSTM model was selected based on the correlation of the input and output crack length. It was determined that a simple LSTM model can reproduce both the noise and general trends. However, it was found that the noise, which includes information about the underlying micro-scale phenomenon requires an input will corresponding high frequency. Therefore, for this simple BiLSTM network the noise in the output is strongly correlated with the input noise. In the absence of this noise in the input, no noise is found in the output. The model proved to be accurate at both interpolating intermediate initial stress intensities and temperatures. The model can produce all intermediate conditions in the absence of conducting more experiments, which provides a tool for producing better damage models. Given the ability of the BiLSTM and the promise shown by it to accurately model the fatigue properties of IN718, the damage model developed using the BiLSTM could be used to prepare for maintenance. With IN718's dependence on the micro-structure of the material, which could include cracks, the use of a model to predict a given crack length could be invaluable to keeping efficiency high. Similarly, since the running of the experiments used to train the model can be lengthy, the model could be used as a complement or supplement to running the experiments, which in turn will save time. 

\section{Data Availability:}
Machine learning model will be available on GitHub and experimental data can be requested from corresponding author.

\section{Acknowledgements:}
The authors would like to recognize the funding support from the National Science Foundation Award \#1709568. The authors would also like to recognize the initial discussion of this manuscript with Stefanos Papanikolaou. T.M. would like to acknowledge the experimental assistance of Lucas Ware and Liam Thomas.\\

\section{Author Contribution:}
J.K.-E. was main developer of ML code, ML result synthesis, and writing of manuscript. A.P. conducted experimental study and provided data for training ML model, and writing of manuscript. R.T. contributed to ML model development, and writing of manuscript. T.M. contributed to ML model development, experimental data synthesis, and writing of manuscript.

\section{Competing Interests:}
The authors declare no competing interests.

\bibliography{apssamp}

\providecommand{\noopsort}[1]{}\providecommand{\singleletter}[1]{#1}%
\begin{thebibliography}{35}%
\makeatletter
\providecommand \@ifxundefined [1]{%
 \@ifx{#1\undefined}
}%
\providecommand \@ifnum [1]{%
 \ifnum #1\expandafter \@firstoftwo
 \else \expandafter \@secondoftwo
 \fi
}%
\providecommand \@ifx [1]{%
 \ifx #1\expandafter \@firstoftwo
 \else \expandafter \@secondoftwo
 \fi
}%
\providecommand \natexlab [1]{#1}%
\providecommand \enquote  [1]{``#1''}%
\providecommand \bibnamefont  [1]{#1}%
\providecommand \bibfnamefont [1]{#1}%
\providecommand \citenamefont [1]{#1}%
\providecommand \href@noop [0]{\@secondoftwo}%
\providecommand \href [0]{\begingroup \@sanitize@url \@href}%
\providecommand \@href[1]{\@@startlink{#1}\@@href}%
\providecommand \@@href[1]{\endgroup#1\@@endlink}%
\providecommand \@sanitize@url [0]{\catcode `\\12\catcode `\$12\catcode
  `\&12\catcode `\#12\catcode `\^12\catcode `\_12\catcode `\%12\relax}%
\providecommand \@@startlink[1]{}%
\providecommand \@@endlink[0]{}%
\providecommand \url  [0]{\begingroup\@sanitize@url \@url }%
\providecommand \@url [1]{\endgroup\@href {#1}{\urlprefix }}%
\providecommand \urlprefix  [0]{URL }%
\providecommand \Eprint [0]{\href }%
\providecommand \doibase [0]{https://doi.org/}%
\providecommand \selectlanguage [0]{\@gobble}%
\providecommand \bibinfo  [0]{\@secondoftwo}%
\providecommand \bibfield  [0]{\@secondoftwo}%
\providecommand \translation [1]{[#1]}%
\providecommand \BibitemOpen [0]{}%
\providecommand \bibitemStop [0]{}%
\providecommand \bibitemNoStop [0]{.\EOS\space}%
\providecommand \EOS [0]{\spacefactor3000\relax}%
\providecommand \BibitemShut  [1]{\csname bibitem#1\endcsname}%
\let\auto@bib@innerbib\@empty
\bibitem [{\citenamefont {Totemeier}\ and\ \citenamefont
  {Tian}(2007)}]{totemeier2007creep}%
  \BibitemOpen
  \bibfield  {author} {\bibinfo {author} {\bibfnamefont {T.~C.}\ \bibnamefont
  {Totemeier}}\ and\ \bibinfo {author} {\bibfnamefont {H.}~\bibnamefont
  {Tian}},\ }\bibfield  {title} {\bibinfo {title} {Creep-fatigue--environment
  interactions in inconel 617},\ }\href@noop {} {\bibfield  {journal} {\bibinfo
   {journal} {Materials Science and Engineering: A}\ }\textbf {\bibinfo
  {volume} {468}},\ \bibinfo {pages} {81} (\bibinfo {year} {2007})}\BibitemShut
  {NoStop}%
\bibitem [{\citenamefont {Kawagoishi}\ \emph {et~al.}(2000)\citenamefont
  {Kawagoishi}, \citenamefont {Chen},\ and\ \citenamefont
  {Nisitani}}]{kawagoishi2000fatigue}%
  \BibitemOpen
  \bibfield  {author} {\bibinfo {author} {\bibfnamefont {N.}~\bibnamefont
  {Kawagoishi}}, \bibinfo {author} {\bibfnamefont {Q.}~\bibnamefont {Chen}},\
  and\ \bibinfo {author} {\bibfnamefont {H.}~\bibnamefont {Nisitani}},\
  }\bibfield  {title} {\bibinfo {title} {Fatigue strength of inconel 718 at
  elevated temperatures},\ }\href@noop {} {\bibfield  {journal} {\bibinfo
  {journal} {Fatigue and Fracture of Engineering Materials and Structures}\
  }\textbf {\bibinfo {volume} {23}},\ \bibinfo {pages} {209} (\bibinfo {year}
  {2000})}\BibitemShut {NoStop}%
\bibitem [{\citenamefont {Gustafsson}\ \emph {et~al.}(2013)\citenamefont
  {Gustafsson}, \citenamefont {Lundstr{\"o}m},\ and\ \citenamefont
  {Simonsson}}]{gustafsson2013modelling}%
  \BibitemOpen
  \bibfield  {author} {\bibinfo {author} {\bibfnamefont {D.}~\bibnamefont
  {Gustafsson}}, \bibinfo {author} {\bibfnamefont {E.}~\bibnamefont
  {Lundstr{\"o}m}},\ and\ \bibinfo {author} {\bibfnamefont {K.}~\bibnamefont
  {Simonsson}},\ }\bibfield  {title} {\bibinfo {title} {Modelling of high
  temperature fatigue crack growth in inconel 718 under hold time conditions},\
  }\href@noop {} {\bibfield  {journal} {\bibinfo  {journal} {International
  journal of fatigue}\ }\textbf {\bibinfo {volume} {52}},\ \bibinfo {pages}
  {124} (\bibinfo {year} {2013})}\BibitemShut {NoStop}%
\bibitem [{\citenamefont {Stephens}\ \emph {et~al.}(2000)\citenamefont
  {Stephens}, \citenamefont {Fatemi}, \citenamefont {Stephens},\ and\
  \citenamefont {Fuchs}}]{stephens2000metal}%
  \BibitemOpen
  \bibfield  {author} {\bibinfo {author} {\bibfnamefont {R.~I.}\ \bibnamefont
  {Stephens}}, \bibinfo {author} {\bibfnamefont {A.}~\bibnamefont {Fatemi}},
  \bibinfo {author} {\bibfnamefont {R.~R.}\ \bibnamefont {Stephens}},\ and\
  \bibinfo {author} {\bibfnamefont {H.~O.}\ \bibnamefont {Fuchs}},\ }\href@noop
  {} {\emph {\bibinfo {title} {Metal fatigue in engineering}}}\ (\bibinfo
  {publisher} {John Wiley \& Sons},\ \bibinfo {year} {2000})\BibitemShut
  {NoStop}%
\bibitem [{\citenamefont {Gustafsson}\ \emph {et~al.}(2011)\citenamefont
  {Gustafsson}, \citenamefont {Moverare}, \citenamefont {Johansson},
  \citenamefont {Simonsson}, \citenamefont {H{\"o}rnqvist}, \citenamefont
  {M{\aa}nsson},\ and\ \citenamefont
  {Sj{\"o}str{\"o}m}}]{gustafsson2011influence}%
  \BibitemOpen
  \bibfield  {author} {\bibinfo {author} {\bibfnamefont {D.}~\bibnamefont
  {Gustafsson}}, \bibinfo {author} {\bibfnamefont {J.}~\bibnamefont
  {Moverare}}, \bibinfo {author} {\bibfnamefont {S.}~\bibnamefont {Johansson}},
  \bibinfo {author} {\bibfnamefont {K.}~\bibnamefont {Simonsson}}, \bibinfo
  {author} {\bibfnamefont {M.}~\bibnamefont {H{\"o}rnqvist}}, \bibinfo {author}
  {\bibfnamefont {T.}~\bibnamefont {M{\aa}nsson}},\ and\ \bibinfo {author}
  {\bibfnamefont {S.}~\bibnamefont {Sj{\"o}str{\"o}m}},\ }\bibfield  {title}
  {\bibinfo {title} {Influence of high temperature hold times on the fatigue
  crack propagation in inconel 718},\ }\href@noop {} {\bibfield  {journal}
  {\bibinfo  {journal} {International Journal of Fatigue}\ }\textbf {\bibinfo
  {volume} {33}},\ \bibinfo {pages} {1461} (\bibinfo {year}
  {2011})}\BibitemShut {NoStop}%
\bibitem [{\citenamefont {Pook}(2007)}]{pook2007metal}%
  \BibitemOpen
  \bibfield  {author} {\bibinfo {author} {\bibfnamefont {L.}~\bibnamefont
  {Pook}},\ }\href@noop {} {\emph {\bibinfo {title} {Why Metal Fatigue
  Matters}}}\ (\bibinfo  {publisher} {Springer},\ \bibinfo {year}
  {2007})\BibitemShut {NoStop}%
\bibitem [{\citenamefont {Ding}\ \emph {et~al.}(2007)\citenamefont {Ding},
  \citenamefont {He}, \citenamefont {Chen}, \citenamefont {Zhu}, \citenamefont
  {Liu},\ and\ \citenamefont {Crepeau}}]{ding2007brief}%
  \BibitemOpen
  \bibfield  {author} {\bibinfo {author} {\bibfnamefont {X.}~\bibnamefont
  {Ding}}, \bibinfo {author} {\bibfnamefont {G.}~\bibnamefont {He}}, \bibinfo
  {author} {\bibfnamefont {C.}~\bibnamefont {Chen}}, \bibinfo {author}
  {\bibfnamefont {Z.}~\bibnamefont {Zhu}}, \bibinfo {author} {\bibfnamefont
  {X.}~\bibnamefont {Liu}},\ and\ \bibinfo {author} {\bibfnamefont {P.~N.}\
  \bibnamefont {Crepeau}},\ }\bibfield  {title} {\bibinfo {title} {A brief
  review of multiaxial high-cycle fatigue},\ }\href@noop {} {\bibfield
  {journal} {\bibinfo  {journal} {Metallurgical and Materials Transactions B}\
  }\textbf {\bibinfo {volume} {38}},\ \bibinfo {pages} {591} (\bibinfo {year}
  {2007})}\BibitemShut {NoStop}%
\bibitem [{\citenamefont {Chan}(2010)}]{chan2010roles}%
  \BibitemOpen
  \bibfield  {author} {\bibinfo {author} {\bibfnamefont {K.~S.}\ \bibnamefont
  {Chan}},\ }\bibfield  {title} {\bibinfo {title} {Roles of microstructure in
  fatigue crack initiation},\ }\href@noop {} {\bibfield  {journal} {\bibinfo
  {journal} {International Journal of Fatigue}\ }\textbf {\bibinfo {volume}
  {32}},\ \bibinfo {pages} {1428} (\bibinfo {year} {2010})}\BibitemShut
  {NoStop}%
\bibitem [{\citenamefont {Biallas}\ and\ \citenamefont
  {Maier}(2007)}]{biallas2007situ}%
  \BibitemOpen
  \bibfield  {author} {\bibinfo {author} {\bibfnamefont {G.}~\bibnamefont
  {Biallas}}\ and\ \bibinfo {author} {\bibfnamefont {H.~J.}\ \bibnamefont
  {Maier}},\ }\bibfield  {title} {\bibinfo {title} {In-situ fatigue in an
  environmental scanning electron microscope--potential and current
  limitations},\ }\href@noop {} {\bibfield  {journal} {\bibinfo  {journal}
  {International journal of fatigue}\ }\textbf {\bibinfo {volume} {29}},\
  \bibinfo {pages} {1413} (\bibinfo {year} {2007})}\BibitemShut {NoStop}%
\bibitem [{\citenamefont {Kamble}\ \emph {et~al.}(2021)\citenamefont {Kamble},
  \citenamefont {Raykar},\ and\ \citenamefont {Jadhav}}]{kamble2021machine}%
  \BibitemOpen
  \bibfield  {author} {\bibinfo {author} {\bibfnamefont {R.~G.}\ \bibnamefont
  {Kamble}}, \bibinfo {author} {\bibfnamefont {N.}~\bibnamefont {Raykar}},\
  and\ \bibinfo {author} {\bibfnamefont {D.}~\bibnamefont {Jadhav}},\
  }\bibfield  {title} {\bibinfo {title} {Machine learning approach to predict
  fatigue crack growth},\ }\href@noop {} {\bibfield  {journal} {\bibinfo
  {journal} {Materials Today: Proceedings}\ }\textbf {\bibinfo {volume} {38}},\
  \bibinfo {pages} {2506} (\bibinfo {year} {2021})}\BibitemShut {NoStop}%
\bibitem [{\citenamefont {Zhan}\ and\ \citenamefont
  {Li}(2021)}]{zhan2021novel}%
  \BibitemOpen
  \bibfield  {author} {\bibinfo {author} {\bibfnamefont {Z.}~\bibnamefont
  {Zhan}}\ and\ \bibinfo {author} {\bibfnamefont {H.}~\bibnamefont {Li}},\
  }\bibfield  {title} {\bibinfo {title} {A novel approach based on the
  elastoplastic fatigue damage and machine learning models for life prediction
  of aerospace alloy parts fabricated by additive manufacturing},\ }\href@noop
  {} {\bibfield  {journal} {\bibinfo  {journal} {International Journal of
  Fatigue}\ }\textbf {\bibinfo {volume} {145}},\ \bibinfo {pages} {106089}
  (\bibinfo {year} {2021})}\BibitemShut {NoStop}%
\bibitem [{\citenamefont {Luo}\ \emph {et~al.}(2021)\citenamefont {Luo},
  \citenamefont {Zhang}, \citenamefont {Feng}, \citenamefont {Song},
  \citenamefont {Qi}, \citenamefont {Li}, \citenamefont {Chen},\ and\
  \citenamefont {Zhang}}]{luo2021pore}%
  \BibitemOpen
  \bibfield  {author} {\bibinfo {author} {\bibfnamefont {Y.}~\bibnamefont
  {Luo}}, \bibinfo {author} {\bibfnamefont {B.}~\bibnamefont {Zhang}}, \bibinfo
  {author} {\bibfnamefont {X.}~\bibnamefont {Feng}}, \bibinfo {author}
  {\bibfnamefont {Z.}~\bibnamefont {Song}}, \bibinfo {author} {\bibfnamefont
  {X.}~\bibnamefont {Qi}}, \bibinfo {author} {\bibfnamefont {C.}~\bibnamefont
  {Li}}, \bibinfo {author} {\bibfnamefont {G.}~\bibnamefont {Chen}},\ and\
  \bibinfo {author} {\bibfnamefont {G.}~\bibnamefont {Zhang}},\ }\bibfield
  {title} {\bibinfo {title} {Pore-affected fatigue life scattering and
  prediction of additively manufactured inconel 718: An investigation based on
  miniature specimen testing and machine learning approach},\ }\href@noop {}
  {\bibfield  {journal} {\bibinfo  {journal} {Materials Science and
  Engineering: A}\ }\textbf {\bibinfo {volume} {802}},\ \bibinfo {pages}
  {140693} (\bibinfo {year} {2021})}\BibitemShut {NoStop}%
\bibitem [{\citenamefont {Sanchez}\ \emph {et~al.}(2021)\citenamefont
  {Sanchez}, \citenamefont {Rengasamy}, \citenamefont {Hyde}, \citenamefont
  {Figueredo},\ and\ \citenamefont {Rothwell}}]{sanchez2021machine}%
  \BibitemOpen
  \bibfield  {author} {\bibinfo {author} {\bibfnamefont {S.}~\bibnamefont
  {Sanchez}}, \bibinfo {author} {\bibfnamefont {D.}~\bibnamefont {Rengasamy}},
  \bibinfo {author} {\bibfnamefont {C.~J.}\ \bibnamefont {Hyde}}, \bibinfo
  {author} {\bibfnamefont {G.~P.}\ \bibnamefont {Figueredo}},\ and\ \bibinfo
  {author} {\bibfnamefont {B.}~\bibnamefont {Rothwell}},\ }\bibfield  {title}
  {\bibinfo {title} {Machine learning to determine the main factors affecting
  creep rates in laser powder bed fusion},\ }\href@noop {} {\bibfield
  {journal} {\bibinfo  {journal} {Journal of Intelligent Manufacturing}\ ,\
  \bibinfo {pages} {1}} (\bibinfo {year} {2021})}\BibitemShut {NoStop}%
\bibitem [{\citenamefont {Liu}\ \emph {et~al.}(2020)\citenamefont {Liu},
  \citenamefont {Wu}, \citenamefont {Wang}, \citenamefont {Lu}, \citenamefont
  {Avdeev}, \citenamefont {Shi}, \citenamefont {Wang},\ and\ \citenamefont
  {Yu}}]{liu2020predicting}%
  \BibitemOpen
  \bibfield  {author} {\bibinfo {author} {\bibfnamefont {Y.}~\bibnamefont
  {Liu}}, \bibinfo {author} {\bibfnamefont {J.}~\bibnamefont {Wu}}, \bibinfo
  {author} {\bibfnamefont {Z.}~\bibnamefont {Wang}}, \bibinfo {author}
  {\bibfnamefont {X.-G.}\ \bibnamefont {Lu}}, \bibinfo {author} {\bibfnamefont
  {M.}~\bibnamefont {Avdeev}}, \bibinfo {author} {\bibfnamefont
  {S.}~\bibnamefont {Shi}}, \bibinfo {author} {\bibfnamefont {C.}~\bibnamefont
  {Wang}},\ and\ \bibinfo {author} {\bibfnamefont {T.}~\bibnamefont {Yu}},\
  }\bibfield  {title} {\bibinfo {title} {Predicting creep rupture life of
  ni-based single crystal superalloys using divide-and-conquer approach based
  machine learning},\ }\href@noop {} {\bibfield  {journal} {\bibinfo  {journal}
  {Acta Materialia}\ }\textbf {\bibinfo {volume} {195}},\ \bibinfo {pages}
  {454} (\bibinfo {year} {2020})}\BibitemShut {NoStop}%
\bibitem [{\citenamefont {Zhang}\ \emph {et~al.}(2021)\citenamefont {Zhang},
  \citenamefont {Gong},\ and\ \citenamefont {Xuan}}]{ZHANG2021106236}%
  \BibitemOpen
  \bibfield  {author} {\bibinfo {author} {\bibfnamefont {X.-C.}\ \bibnamefont
  {Zhang}}, \bibinfo {author} {\bibfnamefont {J.-G.}\ \bibnamefont {Gong}},\
  and\ \bibinfo {author} {\bibfnamefont {F.-Z.}\ \bibnamefont {Xuan}},\
  }\bibfield  {title} {\bibinfo {title} {A deep learning based life prediction
  method for components under creep, fatigue and creep-fatigue conditions},\
  }\href {https://doi.org/https://doi.org/10.1016/j.ijfatigue.2021.106236}
  {\bibfield  {journal} {\bibinfo  {journal} {International Journal of
  Fatigue}\ }\textbf {\bibinfo {volume} {148}},\ \bibinfo {pages} {106236}
  (\bibinfo {year} {2021})}\BibitemShut {NoStop}%
\bibitem [{\citenamefont {Bao}\ \emph {et~al.}(2021)\citenamefont {Bao},
  \citenamefont {Wu}, \citenamefont {Zhengkai}, \citenamefont {Kang},
  \citenamefont {Peng},\ and\ \citenamefont {Withers}}]{mlfatigue}%
  \BibitemOpen
  \bibfield  {author} {\bibinfo {author} {\bibfnamefont {H.}~\bibnamefont
  {Bao}}, \bibinfo {author} {\bibfnamefont {S.}~\bibnamefont {Wu}}, \bibinfo
  {author} {\bibfnamefont {W.}~\bibnamefont {Zhengkai}}, \bibinfo {author}
  {\bibfnamefont {G.}~\bibnamefont {Kang}}, \bibinfo {author} {\bibfnamefont
  {X.}~\bibnamefont {Peng}},\ and\ \bibinfo {author} {\bibfnamefont
  {P.}~\bibnamefont {Withers}},\ }\bibfield  {title} {\bibinfo {title} {A
  machine-learning fatigue life prediction approach of additively manufactured
  metals},\ }\href {https://doi.org/10.1016/j.engfracmech.2020.107508}
  {\bibfield  {journal} {\bibinfo  {journal} {Engineering Fracture Mechanics}\
  }\textbf {\bibinfo {volume} {242}},\ \bibinfo {pages} {107508} (\bibinfo
  {year} {2021})}\BibitemShut {NoStop}%
\bibitem [{\citenamefont {E647}(2021)}]{astm2018dcpd}%
  \BibitemOpen
  \bibfield  {author} {\bibinfo {author} {\bibfnamefont {A.}~\bibnamefont
  {E647}},\ }\bibfield  {title} {\bibinfo {title} {Standard test method for
  measurement of fatigue crack growth rates},\ }\href@noop {} {\bibfield
  {journal} {\bibinfo  {journal} {ASTM International}\ } (\bibinfo {year}
  {2021})}\BibitemShut {NoStop}%
\bibitem [{\citenamefont {Wang}\ \emph {et~al.}(2017)\citenamefont {Wang},
  \citenamefont {Zhang}, \citenamefont {Sun},\ and\ \citenamefont
  {Zhang}}]{wang2017comparison}%
  \BibitemOpen
  \bibfield  {author} {\bibinfo {author} {\bibfnamefont {H.}~\bibnamefont
  {Wang}}, \bibinfo {author} {\bibfnamefont {W.}~\bibnamefont {Zhang}},
  \bibinfo {author} {\bibfnamefont {F.}~\bibnamefont {Sun}},\ and\ \bibinfo
  {author} {\bibfnamefont {W.}~\bibnamefont {Zhang}},\ }\bibfield  {title}
  {\bibinfo {title} {A comparison study of machine learning based algorithms
  for fatigue crack growth calculation},\ }\href@noop {} {\bibfield  {journal}
  {\bibinfo  {journal} {Materials}\ }\textbf {\bibinfo {volume} {10}},\
  \bibinfo {pages} {543} (\bibinfo {year} {2017})}\BibitemShut {NoStop}%
\bibitem [{\citenamefont {Hu}\ \emph {et~al.}(2020)\citenamefont {Hu},
  \citenamefont {Su}, \citenamefont {Liu}, \citenamefont {Mao}, \citenamefont
  {Shan},\ and\ \citenamefont {Wang}}]{hu2020bayesian}%
  \BibitemOpen
  \bibfield  {author} {\bibinfo {author} {\bibfnamefont {D.}~\bibnamefont
  {Hu}}, \bibinfo {author} {\bibfnamefont {X.}~\bibnamefont {Su}}, \bibinfo
  {author} {\bibfnamefont {X.}~\bibnamefont {Liu}}, \bibinfo {author}
  {\bibfnamefont {J.}~\bibnamefont {Mao}}, \bibinfo {author} {\bibfnamefont
  {X.}~\bibnamefont {Shan}},\ and\ \bibinfo {author} {\bibfnamefont
  {R.}~\bibnamefont {Wang}},\ }\bibfield  {title} {\bibinfo {title}
  {Bayesian-based probabilistic fatigue crack growth evaluation combined with
  machine-learning-assisted gpr},\ }\href@noop {} {\bibfield  {journal}
  {\bibinfo  {journal} {Engineering Fracture Mechanics}\ }\textbf {\bibinfo
  {volume} {229}},\ \bibinfo {pages} {106933} (\bibinfo {year}
  {2020})}\BibitemShut {NoStop}%
\bibitem [{\citenamefont {Paris}\ \emph {et~al.}(1999)\citenamefont {Paris},
  \citenamefont {Tada},\ and\ \citenamefont {Donald}}]{paris1999service}%
  \BibitemOpen
  \bibfield  {author} {\bibinfo {author} {\bibfnamefont {P.~C.}\ \bibnamefont
  {Paris}}, \bibinfo {author} {\bibfnamefont {H.}~\bibnamefont {Tada}},\ and\
  \bibinfo {author} {\bibfnamefont {J.~K.}\ \bibnamefont {Donald}},\ }\bibfield
   {title} {\bibinfo {title} {Service load fatigue damage—a historical
  perspective},\ }\href@noop {} {\bibfield  {journal} {\bibinfo  {journal}
  {International Journal of fatigue}\ }\textbf {\bibinfo {volume} {21}},\
  \bibinfo {pages} {S35} (\bibinfo {year} {1999})}\BibitemShut {NoStop}%
\bibitem [{\citenamefont {Rovinelli}\ \emph {et~al.}(2018)\citenamefont
  {Rovinelli}, \citenamefont {Sangid}, \citenamefont {Proudhon},\ and\
  \citenamefont {Ludwig}}]{rovinelli2018using}%
  \BibitemOpen
  \bibfield  {author} {\bibinfo {author} {\bibfnamefont {A.}~\bibnamefont
  {Rovinelli}}, \bibinfo {author} {\bibfnamefont {M.~D.}\ \bibnamefont
  {Sangid}}, \bibinfo {author} {\bibfnamefont {H.}~\bibnamefont {Proudhon}},\
  and\ \bibinfo {author} {\bibfnamefont {W.}~\bibnamefont {Ludwig}},\
  }\bibfield  {title} {\bibinfo {title} {Using machine learning and a
  data-driven approach to identify the small fatigue crack driving force in
  polycrystalline materials},\ }\href@noop {} {\bibfield  {journal} {\bibinfo
  {journal} {npj Computational Materials}\ }\textbf {\bibinfo {volume} {4}},\
  \bibinfo {pages} {1} (\bibinfo {year} {2018})}\BibitemShut {NoStop}%
\bibitem [{\citenamefont {Choi}\ \emph {et~al.}(2021)\citenamefont {Choi},
  \citenamefont {Quagliato}, \citenamefont {Lee}, \citenamefont {Shin},\ and\
  \citenamefont {Kim}}]{choi2021multiaxial}%
  \BibitemOpen
  \bibfield  {author} {\bibinfo {author} {\bibfnamefont {J.}~\bibnamefont
  {Choi}}, \bibinfo {author} {\bibfnamefont {L.}~\bibnamefont {Quagliato}},
  \bibinfo {author} {\bibfnamefont {S.}~\bibnamefont {Lee}}, \bibinfo {author}
  {\bibfnamefont {J.}~\bibnamefont {Shin}},\ and\ \bibinfo {author}
  {\bibfnamefont {N.}~\bibnamefont {Kim}},\ }\bibfield  {title} {\bibinfo
  {title} {Multiaxial fatigue life prediction of polychloroprene rubber (cr)
  reinforced with tungsten nano-particles based on semi-empirical and machine
  learning models},\ }\href@noop {} {\bibfield  {journal} {\bibinfo  {journal}
  {International Journal of Fatigue}\ }\textbf {\bibinfo {volume} {145}},\
  \bibinfo {pages} {106136} (\bibinfo {year} {2021})}\BibitemShut {NoStop}%
\bibitem [{\citenamefont {{Siami-Namini}}\ \emph {et~al.}(2019)\citenamefont
  {{Siami-Namini}}, \citenamefont {{Tavakoli}},\ and\ \citenamefont
  {{Namin}}}]{LSTMvsBLSTM}%
  \BibitemOpen
  \bibfield  {author} {\bibinfo {author} {\bibfnamefont {S.}~\bibnamefont
  {{Siami-Namini}}}, \bibinfo {author} {\bibfnamefont {N.}~\bibnamefont
  {{Tavakoli}}},\ and\ \bibinfo {author} {\bibfnamefont {A.~S.}\ \bibnamefont
  {{Namin}}},\ }\bibfield  {title} {\bibinfo {title} {The performance of lstm
  and bilstm in forecasting time series},\ }in\ \href@noop {} {\emph {\bibinfo
  {booktitle} {2019 IEEE International Conference on Big Data (Big Data)}}}\
  (\bibinfo {year} {2019})\ pp.\ \bibinfo {pages} {3285--3292}\BibitemShut
  {NoStop}%
\bibitem [{\citenamefont {{Althelaya}}\ \emph {et~al.}(2018)\citenamefont
  {{Althelaya}}, \citenamefont {{El-Alfy}},\ and\ \citenamefont
  {{Mohammed}}}]{StockBLSTM}%
  \BibitemOpen
  \bibfield  {author} {\bibinfo {author} {\bibfnamefont {K.~A.}\ \bibnamefont
  {{Althelaya}}}, \bibinfo {author} {\bibfnamefont {E.~M.}\ \bibnamefont
  {{El-Alfy}}},\ and\ \bibinfo {author} {\bibfnamefont {S.}~\bibnamefont
  {{Mohammed}}},\ }\bibfield  {title} {\bibinfo {title} {Evaluation of
  bidirectional lstm for short-and long-term stock market prediction},\ }in\
  \href@noop {} {\emph {\bibinfo {booktitle} {2018 9th International Conference
  on Information and Communication Systems (ICICS)}}}\ (\bibinfo {year}
  {2018})\ pp.\ \bibinfo {pages} {151--156}\BibitemShut {NoStop}%
\bibitem [{\citenamefont {Pokharel}\ \emph {et~al.}(2019)\citenamefont
  {Pokharel}, \citenamefont {Lindsay}, , \citenamefont {Papanikolaou},\ and\
  \citenamefont {Musho}}]{ansan}%
  \BibitemOpen
  \bibfield  {author} {\bibinfo {author} {\bibfnamefont {A.}~\bibnamefont
  {Pokharel}}, \bibinfo {author} {\bibfnamefont {J.}~\bibnamefont {Lindsay}}, ,
  \bibinfo {author} {\bibfnamefont {S.}~\bibnamefont {Papanikolaou}},\ and\
  \bibinfo {author} {\bibfnamefont {T.}~\bibnamefont {Musho}},\ }\bibfield
  {title} {\bibinfo {title} {Experimental investigation of stochastic jumps
  during crack initiation and growth in in718},\ }\href@noop {} {\bibfield
  {journal} {\bibinfo  {journal} {arXiv preprint arXiv:1906.03681}\ } (\bibinfo
  {year} {2019})}\BibitemShut {NoStop}%
\bibitem [{\citenamefont {Hochreiter}\ and\ \citenamefont
  {Schmidhuber}(1997)}]{hochreiter1997long}%
  \BibitemOpen
  \bibfield  {author} {\bibinfo {author} {\bibfnamefont {S.}~\bibnamefont
  {Hochreiter}}\ and\ \bibinfo {author} {\bibfnamefont {J.}~\bibnamefont
  {Schmidhuber}},\ }\bibfield  {title} {\bibinfo {title} {Long short-term
  memory},\ }\href@noop {} {\bibfield  {journal} {\bibinfo  {journal} {Neural
  computation}\ }\textbf {\bibinfo {volume} {9}},\ \bibinfo {pages} {1735}
  (\bibinfo {year} {1997})}\BibitemShut {NoStop}%
\bibitem [{\citenamefont {Trinh}\ \emph {et~al.}(2018)\citenamefont {Trinh},
  \citenamefont {Dai}, \citenamefont {Luong},\ and\ \citenamefont
  {Le}}]{DBLP:journals/corr/abs-1803-00144}%
  \BibitemOpen
  \bibfield  {author} {\bibinfo {author} {\bibfnamefont {T.~H.}\ \bibnamefont
  {Trinh}}, \bibinfo {author} {\bibfnamefont {A.~M.}\ \bibnamefont {Dai}},
  \bibinfo {author} {\bibfnamefont {T.}~\bibnamefont {Luong}},\ and\ \bibinfo
  {author} {\bibfnamefont {Q.~V.}\ \bibnamefont {Le}},\ }\bibfield  {title}
  {\bibinfo {title} {Learning longer-term dependencies in rnns with auxiliary
  losses},\ }\href {http://arxiv.org/abs/1803.00144} {\bibfield  {journal}
  {\bibinfo  {journal} {CoRR}\ }\textbf {\bibinfo {volume} {abs/1803.00144}}
  (\bibinfo {year} {2018})},\ \Eprint {https://arxiv.org/abs/1803.00144}
  {arXiv:1803.00144} \BibitemShut {NoStop}%
\bibitem [{\citenamefont {Gers}\ \emph {et~al.}(2000)\citenamefont {Gers},
  \citenamefont {Schmidhuber},\ and\ \citenamefont {Cummins}}]{vangrad}%
  \BibitemOpen
  \bibfield  {author} {\bibinfo {author} {\bibfnamefont {F.}~\bibnamefont
  {Gers}}, \bibinfo {author} {\bibfnamefont {J.}~\bibnamefont {Schmidhuber}},\
  and\ \bibinfo {author} {\bibfnamefont {F.}~\bibnamefont {Cummins}},\
  }\bibfield  {title} {\bibinfo {title} {Learning to forget: Continual
  prediction with lstm},\ }\href
  {https://www.mitpressjournals.org/doi/pdf/10.1162/089976600300015015}
  {\bibfield  {journal} {\bibinfo  {journal} {Neural Computation}\ }\textbf
  {\bibinfo {volume} {12}},\ \bibinfo {pages} {2451} (\bibinfo {year}
  {2000})}\BibitemShut {NoStop}%
\bibitem [{\citenamefont {{Schuster}}\ and\ \citenamefont
  {{Paliwal}}(1997)}]{Bidirectional}%
  \BibitemOpen
  \bibfield  {author} {\bibinfo {author} {\bibfnamefont {M.}~\bibnamefont
  {{Schuster}}}\ and\ \bibinfo {author} {\bibfnamefont {K.~K.}\ \bibnamefont
  {{Paliwal}}},\ }\bibfield  {title} {\bibinfo {title} {Bidirectional recurrent
  neural networks},\ }\href@noop {} {\bibfield  {journal} {\bibinfo  {journal}
  {IEEE Transactions on Signal Processing}\ }\textbf {\bibinfo {volume} {45}},\
  \bibinfo {pages} {2673} (\bibinfo {year} {1997})}\BibitemShut {NoStop}%
\bibitem [{\citenamefont {E1457-15}(2015)}]{astm2015standard}%
  \BibitemOpen
  \bibfield  {author} {\bibinfo {author} {\bibfnamefont {A.}~\bibnamefont
  {E1457-15}},\ }\bibfield  {title} {\bibinfo {title} {Standard test method for
  measurement of creep crack growth times in metals},\ }\href@noop {}
  {\bibfield  {journal} {\bibinfo  {journal} {ASTM International}\ } (\bibinfo
  {year} {2015})}\BibitemShut {NoStop}%
\bibitem [{\citenamefont {Metals}(2007)}]{Inconel718}%
  \BibitemOpen
  \bibfield  {author} {\bibinfo {author} {\bibfnamefont {S.}~\bibnamefont
  {Metals}},\ }\href@noop {} {\bibinfo {title} {Inconel alloy 718}},\ \bibinfo
  {howpublished} {Special Metals},\ \bibinfo {address} {New Hartford, New York}
  (\bibinfo {year} {2007}),\ \bibinfo {note} {information for Inconel
  718}\BibitemShut {NoStop}%
\bibitem [{\citenamefont {B670-07}(2021)}]{astm2018standard}%
  \BibitemOpen
  \bibfield  {author} {\bibinfo {author} {\bibfnamefont {A.}~\bibnamefont
  {B670-07}},\ }\bibfield  {title} {\bibinfo {title} {Standard specification
  for precipitation-hardening nickel alloy (uns n07718) plate, sheet, and strip
  for high-temperature service},\ }\href@noop {} {\bibfield  {journal}
  {\bibinfo  {journal} {ASTM International}\ } (\bibinfo {year}
  {2021})}\BibitemShut {NoStop}%
\bibitem [{\citenamefont {Merah}(2003)}]{DCPDacc}%
  \BibitemOpen
  \bibfield  {author} {\bibinfo {author} {\bibfnamefont {N.}~\bibnamefont
  {Merah}},\ }\bibfield  {title} {\bibinfo {title} {Detecting and measuring
  flaws using electric potential techniques},\ }\href@noop {} {\bibfield
  {journal} {\bibinfo  {journal} {Journal of Quality in Maintenance
  Engineering}\ }\textbf {\bibinfo {volume} {9}},\ \bibinfo {pages} {160}
  (\bibinfo {year} {2003})}\BibitemShut {NoStop}%
\bibitem [{\citenamefont {Johnson}(1965)}]{johnson1965calibrating}%
  \BibitemOpen
  \bibfield  {author} {\bibinfo {author} {\bibfnamefont {H.}~\bibnamefont
  {Johnson}},\ }\bibfield  {title} {\bibinfo {title} {Calibrating the electric
  potential method for studying slow crack growth},\ }\href@noop {} {\bibfield
  {journal} {\bibinfo  {journal} {Materials Research and Standards}\ }\textbf
  {\bibinfo {volume} {5}},\ \bibinfo {pages} {442} (\bibinfo {year}
  {1965})}\BibitemShut {NoStop}%
\bibitem [{\citenamefont {Wang}\ and\ \citenamefont
  {Schwalbe}(1993)}]{WANG19933}%
  \BibitemOpen
  \bibfield  {author} {\bibinfo {author} {\bibfnamefont {G.-X.}\ \bibnamefont
  {Wang}}\ and\ \bibinfo {author} {\bibfnamefont {K.-H.}\ \bibnamefont
  {Schwalbe}},\ }\bibfield  {title} {\bibinfo {title} {A study of the
  transition from intercrystalline to transcrystalline fatigue crack
  propagation in different ageing conditions of the alloy cu-35\%ni-3.5\%cr},\
  }\href@noop {} {\bibfield  {journal} {\bibinfo  {journal} {International
  Journal of Fatigue}\ }\textbf {\bibinfo {volume} {15}},\ \bibinfo {pages} {3}
  (\bibinfo {year} {1993})}\BibitemShut {NoStop}%
\end{thebibliography}%
\end{document}